\documentclass[twocolumn,prl]{revtex4}

 \pdfoutput=1

\usepackage{graphicx}
\usepackage{color}
\usepackage{epsfig}
\usepackage{subfigure}
\usepackage{natbib}
\usepackage{array}
\usepackage{amsmath}

\begin{document}
\newcolumntype{M}{>{$}c<{$}}

%

\title{Fingerprinting quantum emitters in hexagonal boron nitride using strain} 

\author{Pratibha Dev}
\email{pratibha.dev@howard.edu}
\affiliation{Department of Physics and Astronomy, Howard University, Washington, D.C. 20059, USA}

%
%

\begin{abstract}
Two dimensional van der Waals crystals and their heterostructures provide an exciting alternative to bulk wide bandgap semiconductors as hosts of single photon emitters. Amongst different layered materials, bright and robust defect-based single photon emitters have been observed within hexagonal boron nitride, a layered wide-bandgap semiconductor. Despite research efforts to date, the identities of the deep defects responsible for quantum emissions in hexagonal boron nitride remain unknown. In this theoretical work, I demonstrate that the strain-induced changes in emission frequencies depend on: (i) the detailed nature of the defect states involved in the optical excitations, and (ii) the rich boron chemistry that results in complex interactions between boron atoms. As each defect shows a distinct response to the strain, it can be used not only to tune emission frequencies, but also to identify the quantum emitters in hexagonal boron nitride.
\end{abstract}
\maketitle


The discovery of quantum emitters in different two-dimensional (2D) layered structures is a significant development in the search for qubit-candidates for quantum technologies.~\cite{Imamoglu_WSe2_QE_2015, Pan_WSe2_QE_2015, Vamivakas_WSe2_QE_2015, Potemski_WSe2_QE_2015, Tran1_Nature_2015, Tran2_ACS_Nano2016, Aharonovich_hBN_VNCB_2016, Single_Photon_hBN_2016, Englund_strainTunable_Nature_2017, Loncar_TMD_Nature_2017, LeeBassett_2017, Tran3_acsphotonics_2018} These  van der Waals crystals and their heterostructures provide an exciting alternative to quantum emitters (QEs) within bulk wide bandgap semiconductors.  The surface-only structure of the host 2D materials allows for a facile manipulation of the electronic-structure properties of the 2D layers, and hence the QEs within them, via different means, such as changes in composition at the atomistic level,~\cite{Cress_ion_implant_2016, Aharonovich_hBN_VNCB_2016} creation of heterostructures of 2D layers, ~\cite{Dean_BN_Gr_hetero_2010, Geim_2D_hetero_2013, Hone_hetero_2015} or through application of strain.~\cite{Loncar_TMD_Nature_2017, Englund_strainTunable_Nature_2017}

To date, the nature of quantum emitters in layered materials has remained unclear. Even when it is determined that the quantum emitter is a deep defect, as in the case of hexagonal boron nitride (hBN),~\cite{Tran1_Nature_2015, Tran2_ACS_Nano2016, Aharonovich_hBN_VNCB_2016, Single_Photon_hBN_2016, Englund_strainTunable_Nature_2017, LeeBassett_2017, Tran3_acsphotonics_2018} the exact identity remains unknown. In the case of hBN, experimental attempts to identify the defects are confounded by the widely varying properties of the QEs such as their brightness, emission frequencies and polarizations.~\cite{Tran2_ACS_Nano2016, Menon_ACS_Photonics_2016, Jungwirth_ACS_NanoLett_2016, LeeBassett_2017, Englund_strainTunable_Nature_2017, Vogl_Doherty_2019} These variations are observed not only in different samples, which may be prepared using very different treatments, but also within a sample.  
The emission frequencies of QEs in hBN range from ultraviolet to near-infrared~\cite{Tran1_Nature_2015, Tran2_ACS_Nano2016, Englund_strainTunable_Nature_2017, Vogl_Campbell_acsphotonics_2018, Vogl_Doherty_2019}. The emitters are often grouped according to their zero phonon line (ZPL) energies and phonon sideband shapes~\cite{Tran2_ACS_Nano2016}.
Different groups of emitters possibly correspond to chemically-different defects, or defects with different charge states, or a combination of both. Even for what appears to be the same defect in a sample, a large spectral distribution in the ZPL frequencies is reported~\cite{Tran2_ACS_Nano2016, Englund_strainTunable_Nature_2017}. This distribution is usually attributed to different local strains around a QE species. Hence, understanding how strain affects properties of QEs is an important step towards fully characterizing them. Only a few experiments exist that have directly studied the effect of strain on ZPL frequencies of QEs in hBN~\cite{Englund_strainTunable_Nature_2017, Lazic_strain_hBN_2019}. On the theoretical side, an understanding of the response to strain for different defects has been mostly missing. 



In this work, I address these issues by studying the effects of strain on the excited state properties of different candidate defects within hBN, and show how the response of the defects to the strain can be used to identify the defects themselves. Strain displaces the ions, affecting the orbital degrees of freedom of point-like defects. Energies of the defect states---molecular orbitals constructed from the dangling bonds at the defect site---are modified by the strain, tuning the excitation energies according to: $H_{strain}=\sum_{\Gamma}\, \Lambda_{\Gamma}\,\epsilon_{\Gamma}$, where the strain Hamiltonian, $H_{strain}$, has been projected onto the irreducible representations ($\Gamma$) of the defect's symmetry-group, $\Lambda_{\Gamma}$ are the orbital operators and $\epsilon_{\Gamma}$ are the strain tensors. In order to keep the study focussed, I concentrate on the symmetry-preserving hydrostatic strain (dilation and contraction), which belongs to the most symmetric representation of the group, and merely shifts the energies of the defect states.
As the make-up of the defect states is unique for each defect, the strain-induced modification in defect properties varies from defect to defect. Another important consideration is that under-coordinated boron atoms are involved in the defects. The boron atom occupies a special position in the periodic table due to its highly unconventional chemistry, with a tendency to form multi-center bonds to overcome electronic frustration. I show that the rich chemistry of boron plays a role in the equilibrium geometries adopted by the defects in their ground- and excited-states. In turn, this influences the effects of strain on the ZPL. 



\section*{Results}

   \begin{figure*}[t]
  \begin{center}
   \includegraphics[width=0.8\textwidth]{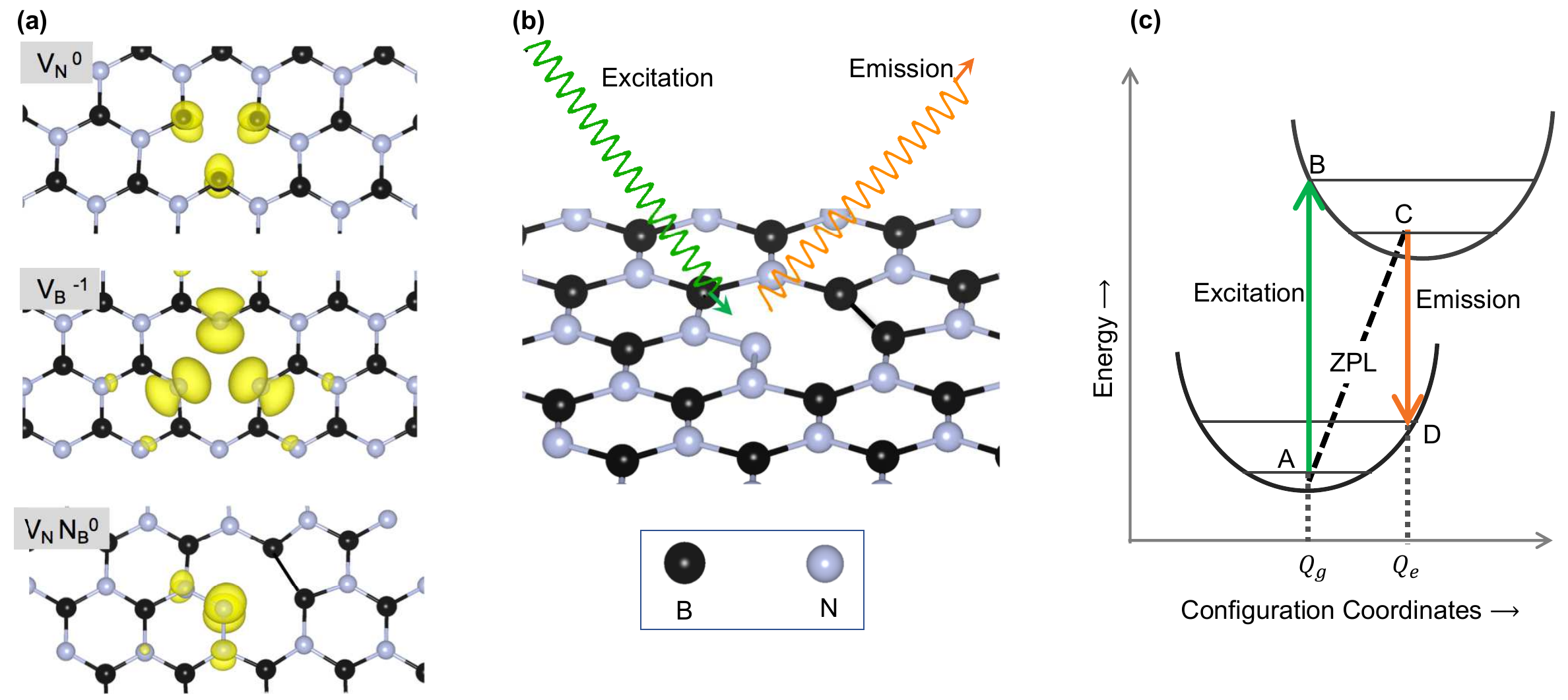}
 \end{center}
 \vspace{-12pt}
\caption{\label{Fig1} Deep defects in hBN as quantum emitters. \textbf{(a)} Spin density plot [difference in charge densities within two spin channels] for defects in unstrained hBN, showing the localized nature of the defect states. \textbf{(b)} Schematic of hBN lattice with a defect undergoing photoluminescence (PL) process. \textbf{(c)} Franck Condon picture used to mimic PL within constrained-occupation DFT (CDFT). The configuration coordinate diagram is a plot of total energy as a function of ground (bottom) and excited state (top) configurations. Within CDFT, total energies corresponding to points A, B, C, and D  are calculated by promoting an electron from a filled defect state to a higher (formerly empty) state to mimic the PL process. 
}
\end{figure*} 

The Quantum-ESPRESSO package~\cite{QE-2009} was used to carry out Density Functional Theory (DFT) calculations within the generalized gradient approximation (GGA)~\cite{GGA} of Perdew-Burke-Ernzerhof (PBE)~\cite{PBE} to approximate the exchange-correlation (xc) energy functional (see Methods). The PBE functional was used even though it is known to underestimate the bandgap~\cite{PBE_bandgap_issue}. This bandgap problem can be resolved by using a computationally expensive hybrid functional such as HSE06~\cite{HSE03, HSE06}, which includes a fixed percentage of Hartree-Fock exchange in the xc-functional. However, in this work, the emphasis is on: (i) identifying the deep defect states through their spatial localization, (ii) understanding how the nature of the optically-active defect states and the involvement of boron atoms play key roles in the response of different defects to the strain, and hence (iii) explaining the trends/changes in the emission frequencies (instead of absolute numbers) as a function of applied strain. As I am mostly interested in aspects of the electronic structure properties of the defects, which are expected to remain the same for the two functionals, I used PBE rather than the prohibitively expensive HSE06 approximation. 

Out of all possible spin-active defects,~\cite{Ford_Tran_2017} I have selected three intrinsic defects of hBN: a neutral nitrogen vacancy ($\mathrm{V_{N}^{\,\,0}}$), a negatively charged boron vacancy ($\mathrm{V_{B}^{\,\,-1}}$), and a neutral antisite complex comprised of a nitrogen vacancy next to a nitrogen substitutional ($\mathrm{V_{N}N_{B}^{\,\,0}}$). The purpose of choosing these particular defects is to demonstrate proof of principle.  The relatively large formation energies~\cite{Weston_PRB_hBN_QE_2018} of two of the defects ($\mathrm{V_{N}^{\,\,0}}$ and $\mathrm{V_{N}N_{B}^{\,\,0}}$) would likely preclude them from existing in large concentrations naturally.  The possibility of observing them increases markedly, however, in systems where the defects are formed by irradiation, a common technique for forming defects in experiments.~\cite{Aharonovich_hBN_VNCB_2016} Through these defects, I demonstrate that the strain not only tunes excitation energies, but it can also be used to characterize defects through their response to the applied strain, which along with other pieces of information, such as symmetry and/or spin, can be used to identify the defects.

The two simple point defects studied in this work, $\mathrm{V_{N}^{\,\,0}}$ and $\mathrm{V_{B}^{\,\,-1}}$, have the same symmetry ($\mathrm{D_{3h}}$), although they have different spins. On the other hand, $\mathrm{V_{N}^{\,\,0}}$  and $\mathrm{V_{N}N_{B}^{\,\,0}}$ have the same spin, although they have different symmetries. 
Each of these defects chosen in this study has partially-filled defect states derived from the dangling $sp^{2}$-hybridized bonds. The spatial localization of the defect states [see Fig.\ref{Fig1}(a)] results in a large exchange interaction and hence, in spin-polarized structures with more electrons in one spin channel (majority spin) than the other (minority-spin channel).  I concentrate on the effect of strain on the lowest-energy excitations for each of the defects. I calculate the ZPL for each of the defects under different levels of strain using the constrained-occupation DFT (CDFT) method, which mimics the photoluminescence process shown in Fig.\ref{Fig1}(b). The total energy differences between different electronic and ionic configurations are further mapped onto the Franck-Condon picture [Fig.\ref{Fig1}(c)], giving optical transition energies.~\cite{Gali_DeltaSCF}

  \begin{figure*}[t]
  \begin{center}
   \includegraphics[width=0.95\textwidth]{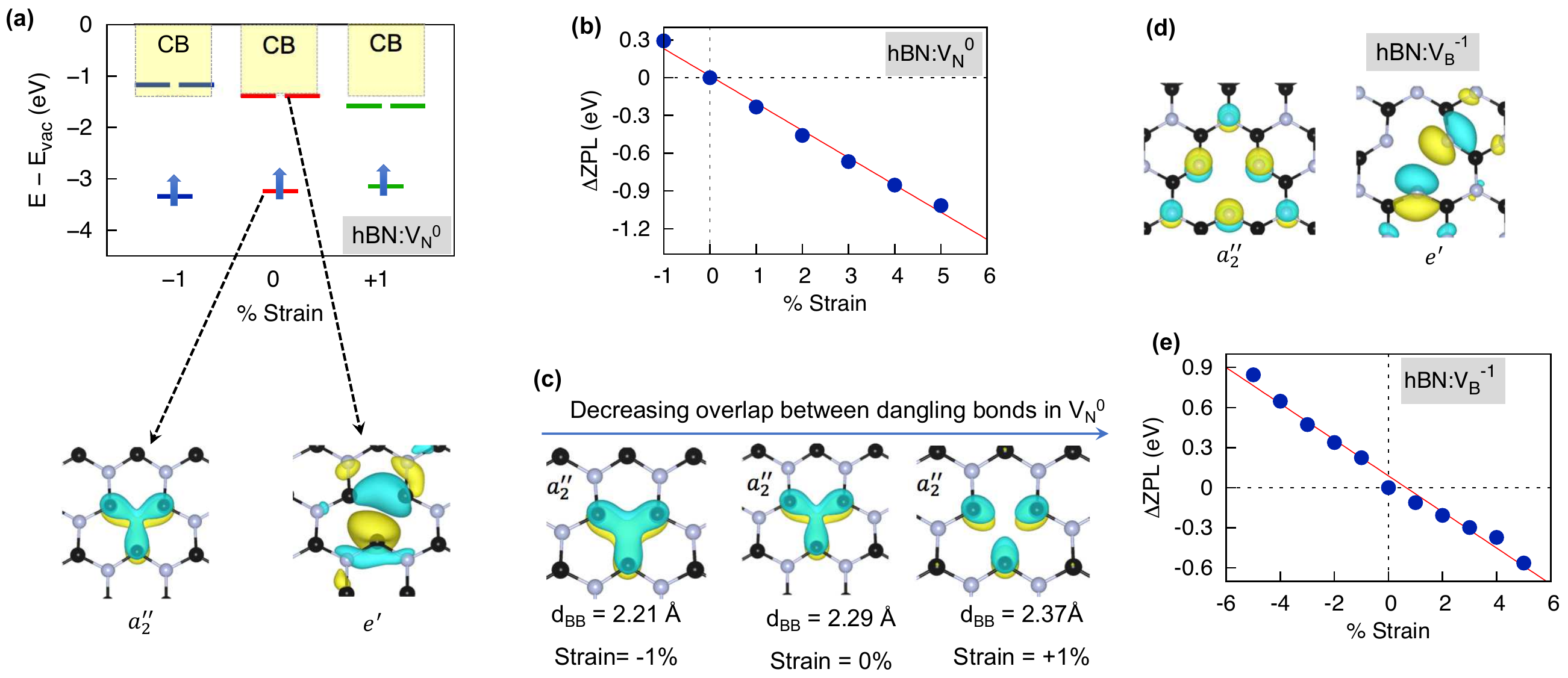}
 \end{center}
 \vspace{-12pt}
 \caption{\label{Fig2} Strain-tuned excitation energies of simple point defects. \textbf{(a)} Single-particle energy levels of the defect states involved in the optical excitation in V$_{\mathrm{N}}^{0}$ as a function of strain, with the vacuum energy used as the reference energy. Optical excitation corresponds to promotion of electron from a filled $a_{2}''$-state (singlet) to the empty $e'$-states (doublet) in the spin-up channel. CB here stands for conduction band. Also shown are the charge density plots for the singlet and one of the doublets. \textbf{(b)} Change in the ZPL  of V$_{\mathrm{N}}^{0}$ as a function of strain (ZPL at 0\% strain used as reference), with the red line being the best fit. \textbf{(c)}  Decreasing overlap between the dangling bonds with the expansion of the lattice (only $a_{2}''$-state shown), increases (decreases) the energy of bonding-like (antibonding) orbitals. Here, $\mathrm{d_{BB}}$ is the distance between boron atoms around the defect. \textbf{(d)} Optically-active defect states involved in the excitation in V$_{\mathrm{B}}^{-1}$. Excitation takes the electron from the filled $a_{2}''$-state (singlet) to the empty $e'$ (doublet) states (only one of the $e'$ states shown). \textbf{(e)} Changes in the ZPL  of V$_{\mathrm{B}}^{-1}$ as a function of strain, with the red line being the best linear fit. 
}
\end{figure*} 

%


\paragraph*{\textbf{Neutral nitrogen vacancy:}} V$_{\mathrm{N}}^{0}$ is a spin-1/2 defect with a $\mathrm{D_{3h}}$ point group symmetry. At 0\% strain the under-coordinated boron atoms surrounding the defect maximize their interactions by moving inwards towards the defect (boron-boron distance, $\mathrm{d_{BB}}=2.29$~\AA\ as compared to 2.51~\AA\ in an ideal crystal). Fig.~\ref{Fig2}(a) shows the optically-active majority spin states for $\mathrm{V_{N}^{0}}$ corresponding to the filled $a_{2}''$-state (singlet) and the empty $e'$-state (doublet) for three different strains. Charge density plots of these states in Fig.~\ref{Fig2}(a) show the bonding character of the $a_{2}''$-state, and the anti-bonding character of the $e'$-state. The strain \% is defined as $(a-a_{0})/a_{0}$, with $a_{0}$ being the equilibrium lattice constant. 
Dilation (contraction) of the lattice results in narrowing (widening) of the gap between the $a_{2}''$-state and the empty $e'$-state. Beyond 1\% compressive strain, the empty $e'$-defect state hybridizes with the bulk states. In experiments, this will appear as a loss of signal from the quantum emitter. These strain-induced changes in separation of optically active levels, result in corresponding changes in the ZPL.  
Fig.~\ref{Fig2}(b) shows that the ZPL is linearly dependent on the applied strain, changing at the rate of $\sim\mathrm{0.216~eV/\%\,strain}$. This can be understood within Molecular Orbital Theory. As seen in the isosurface plots for $a_{2}''$-states in Fig.~\ref{Fig2}(c), lattice expansion increases the distance between boron atoms surrounding the defect. This decreases the overlap between their dangling bonds that make up the molecular orbitals. In turn, this reduction in overlap increases the energy of $a_{2}''$, which is mostly bonding in character, while the energy of the $e'$-states, which are antibonding in character, simultaneously decreases. As a result, the energy difference between the states involved in the excitation is reduced, lowering the ZPL values when tensile strain is applied. 


 %




\paragraph*{\textbf{Negatively-charged boron vacancy:}} V$_{\mathrm{B}}^{-1}$ is a spin-1 defect with a $\mathrm{D_{3h}}$ point group symmetry. The smallest-energy excitation takes place between a filled $a_{2}''$-state (singlet) and an empty $e'$-state (doublet) in the minority spin channel [Fig.~\ref{Fig2}(d)]. The dependence of the ZPL on strain shows monotonic behaviour as seen in Fig.~\ref{Fig2}(e). The energy of quantum emission from the defect decreases (increases) upon lattice dilation (contraction) at the rate of $\sim\mathrm{0.138~eV/\%\,strain}$. The filled $a_{2}''$-state is derived from the dangling $\pi$-orbitals associated with the nearest neighbor (NN) nitrogen atoms and the next-to-nearest neighbour (NNN) nitrogen atoms surrounding the defect [Fig.\ref{Fig2}(d)]. The $a_{2}''$-state is partially bonding (NN nitrogens) and partially antibonding (between NN and NNN nitrogens) in character. As a result, lattice compression and dilation have a smaller effect on the energy of this level. On the other hand, the empty $e'$-state is mostly derived from the dangling $\sigma$-orbitals associated with NN nitrogen atoms and is chiefly antibonding in character.  As a consequence, lattice dilation (compression) lowers (increases) its energy.  This also explains the different rate at which the ZPL of $\mathrm{V_{B}^{\,\,-1}}$ and  $\mathrm{V_{N}^{0}}$ are changed upon application of strain. For $\mathrm{V_{N}^{0}}$, which shows an enhanced response to the strain, the optically-active states have opposite character (bonding vs antibonding), and both states simultaneously change their energies in the opposite sense upon application of strain. There is some contribution to the $e'$-states of $\mathrm{V_{B}^{\,\,-1}}$ coming from the boron atoms as well [Fig.\ref{Fig2}(d)], and this small bonding-character introduced by these boron atoms plays a role as will be discussed later.


   \begin{figure*}[ht]
  \begin{center}
   \includegraphics[width=0.95\textwidth]{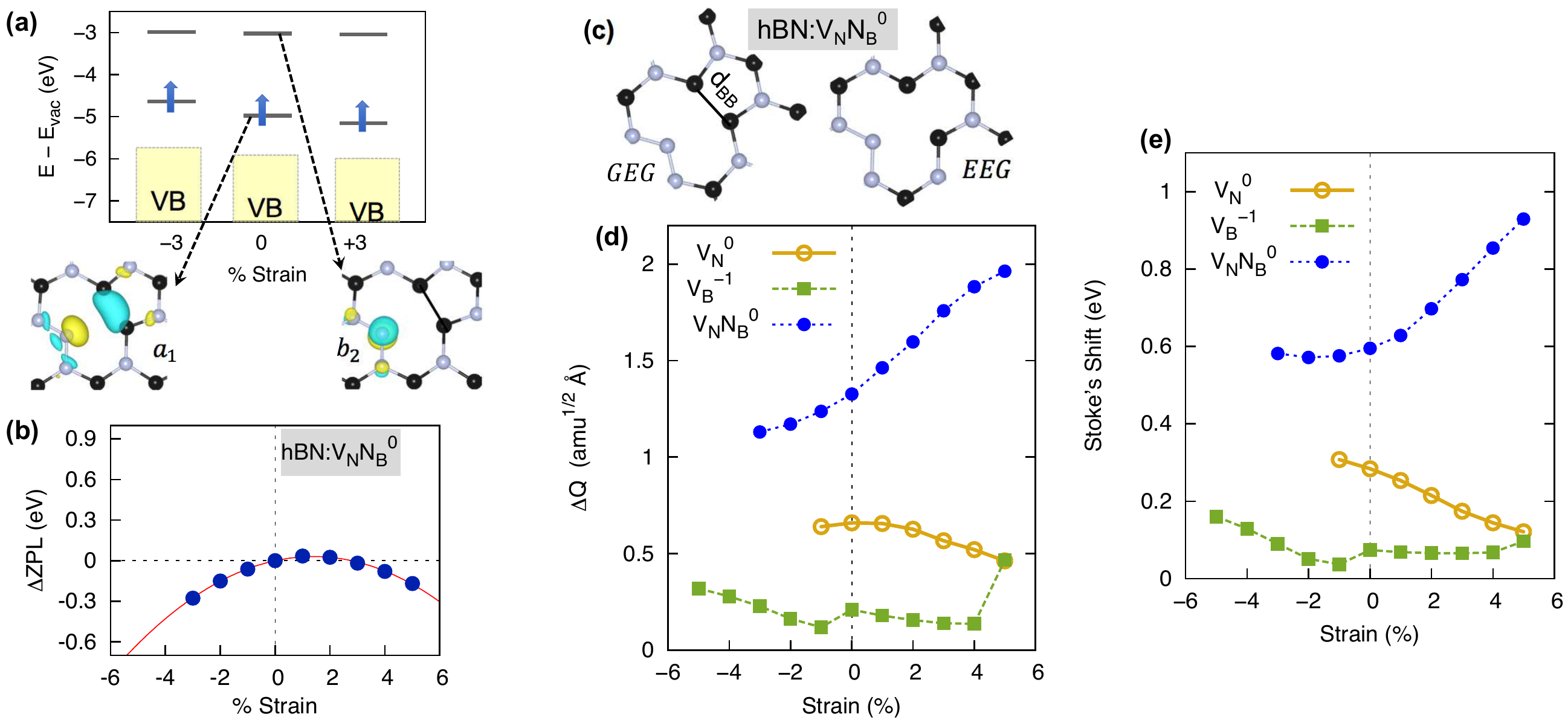}
 \end{center}
 \vspace{-12pt}
 \caption{\label{Fig3}  Distinctive response to strain of the antisite complex. \textbf{(a)} Single-particle energy levels of the defect states involved in the optical excitation in $\mathrm{V_{N}N_{B}^{0}}$ as a function of strain, with the vacuum energy used as the reference energy.  Optical excitation corresponds to promotion of electron from a filled $a_{1}$-state (singlet) to the empty $b_{2}$-state (singlet) in the minority-spin channel. VB here stands for valence band. \textbf{(b)} Change in the ZPL  of $\mathrm{V_{N}N_{B}^{0}}$ as a function of strain (with ZPL at 0\% strain as reference), with the red line being the best fit, showing the non-monotonic behaviour of the ZPL. \textbf{(c)} The ground-state equilibrium geometry (GEG) and the excited-state equilibrium geometry (EEG) of $\mathrm{V_{N}N_{B}^{\,\,0}}$, showing the change in distances ($\mathrm{d_{BB}}$) between the weakly bonded boron atoms. \textbf{(d)} The change in geometry between ground and excited states is quantified by the change in generalized configuration coordinate, $\Delta \mathrm{Q}$, and plotted as a function of strain.  \textbf{(e)} Strain-tuning of the Stoke's shift (difference between excitation energy and the ZPL).
}
\end{figure*} 

\paragraph*{\textbf{Neutral antisite complex:}} $\mathrm{V_{N}N_{B}^{\,\,0}}$ is a spin-1/2 defect, with a $\mathrm{C_{2V}}$ point-group symmetry. The antisite complex is considered to be one of the possible defects that emits in the visible range.~\cite{Tran2_ACS_Nano2016,Englund_strainTunable_Nature_2017,Weston_PRB_hBN_QE_2018,Lazic_strain_hBN_2019} In the ground state equilibrium geometry (GEG), the two boron atoms surrounding $\mathrm{V_{N}N_{B}^{0}}$ form a weak covalent bond, with $\mathrm{d_{BB}}=1.94$~\AA\ for $0\%$ strain. The lowest energy excitation corresponds to the promotion of an electron from a filled singlet ($a_{1}$) state to an empty singlet ($b_{2}$) state in the minority-spin channel. Fig.~\ref{Fig3}(a) gives the defect levels involved in the excitation for different strains, along with their charge density plots. The $a_{1}$-state, which shows antibonding character between the dominant contributing centers ($\mathrm{N_{B}}$ and the two bonded borons), is stabilized (destabilized) upon dilatation (contraction).  On the other hand, the $b_{2}$ state, for which the charge density is mostly localized on the antisite nitrogen atom, is minimally affected by strain. Fig.~\ref{Fig3}(b) shows changes in the ZPL as a function of strain. The ZPL has a non-monotonic behavior, which is unlike the trends shown by $\mathrm{V_{B}^{\,\,-1}}$ and $\mathrm{V_{N}^{0}}$, distinguishing this defect from other intrinsic point defects. This is also different from what one might expect from the increasing energy differences in the $a_{1}$- and the  $b_{2}$-states with lattice expansion [see Fig.~\ref{Fig3}(a)]. Interestingly, our theoretically predicted non-monotonic behavior of ZPL agrees with the results obtained experimentally by G. Grosso, \textit{et al.}.~\cite{Englund_strainTunable_Nature_2017} In this work, the authors suspected that the antisite-complex is their defect, and studied the strain-dependence of its ZPL.


 \paragraph*{\textbf{Discussion:}} The unexpected trend in the strain-tuned ZPL for $\mathrm{V_{N}N_{B}^{0}}$ comes from the rich chemistry of the boron atoms that are involved in the defect. Changes in electronic configuration upon excitation dictate the changes in atomic structure in the excited state. In turn, the strain affects the extent of change in the atomic structures due to excitation. As seen in Fig.~\ref{Fig3}(c), the excited-state equilibrium geometry (EEG) for $\mathrm{V_{N}N_{B}^{0}}$ shows a large change in boron-boron distance ($\mathrm{d_{BB}}=2.28$~\AA). The difference in GEG and EEG can be quantified by calculating the change in the generalized configuration coordinate, $\Delta \mathrm{Q}$, given by the formula: $\Delta \mathrm{Q}^{2}=\sum_{i\alpha}\,m_{\alpha}(R_{i\alpha}^{EEG} - R_{i\alpha}^{GEG})$. Here, $m_{\alpha}$ is the atomic mass of the $\alpha^{th}$ atom and $(R_{i\alpha}^{EEG} - R_{i\alpha}^{GEG})$ is its displacement in the $i^{th}$-direction ($i={x,y,z}$). In the case of lattice contraction, the changes in the geometry of $\mathrm{V_{N}N_{B}^{0}}$ are small due to the strengthening of the boron-boron bond, while lattice dilation leads to large changes in the geometry due to the weakening of the boron-boron bond [Fig.~\ref{Fig3}(d)]. This is different than the strain-dependence of $\Delta \mathrm{Q}$ for the two simple point defects. In the case of $\mathrm{V_{N}}^{0}$, the distance between neighbouring boron atoms changes from $2.29$~\AA\ in ground state to $2.47$~\AA\ in the excited state for $0\%$ strain. Since these atoms are not bonded, there is a much smaller change in $\Delta \mathrm{Q}$ with strain as compared to the antisite complex. This is also true of $\mathrm{V_{B}}^{-1}$, in which case an even smaller role is played by the interactions between the non-bonded boron atoms that are next-to-nearest neighbouring atoms surrounding the defect. The small, but non-negligible contribution of the boron atoms to the $e'$ (doublet) states involved in the excitation [Fig.~\ref{Fig2}(d)] is responsible for the small deviations from the linear fit for the strain-tuning of the ZPL [Fig.~\ref{Fig2}(e)], and is reflected in the strain-dependence of $\Delta \mathrm{Q}$ for $\mathrm{V_{B}}^{-1}$. Fig.~\ref{Fig3}(e) shows the strain-dependent Stoke's shift (difference in the vertical excitation and the ZPL) for the three defects. The much larger change in the strain-tuned Stoke's shift for $\mathrm{V_{N}N_{B}^{0}}$ as compared to the two other defects further supports the above analysis. It also explains the unexpected decrease in the ZPL of $\mathrm{V_{N}N_{B}^{0}}$ with lattice dilation [Fig.~\ref{Fig3}(b)]. This non-monotonic behavior of the ZPL is due to the large increase in Stoke's shift with increasing dilation. 
 
\vspace{6pt}

 \section*{Conclusion} In summary, I demonstrated that the response of the defects to strain depends on the detailed nature of the defect states involved in the optical excitations, and the rich boron chemistry, which results in complex interactions between the boron atoms. Strain not only tunes emission frequencies, but also provides a possible way of finger-printing defects responsible for quantum emission. Identifying defect(s) and better understanding their characteristic response to external stimuli such as strain, in turn, will help to harness their potential in the fields of quantum-information, -sensing, and -photonics.

 

\section*{Methods}

The Density Functional Theory-based calculations were carried out using the Quantum-ESPRESSO package.~\cite{QE-2009} I used the generalized gradient approximation (GGA)~\cite{GGA} of Perdew-Burke-Ernzerhof (PBE).~\cite{PBE} Defects were created in a $6\times6\times1$ (72-atoms) supercell of hBN.  The Brillouin-zone was sampled with a grid of $\Gamma$-centered, $4\times4\times1$ k-points according to the Monkhort-Pack method.~\cite{kpoint2} These calculations yielded the single particle molecular orbitals (MOs) of different defect centers in the layered material. The ordering of the defect states in each of the deep defects was obtained from the calculated Kohn-Sham eigenstates around the bandgap.  After determining the ground state properties for the unstrained structures, I applied compressive and tensile strain to the system that ranged from $-5\%$ to $+5\%$ (unless the system ceased to behave like a QE). The strain \% is defined as $(a-a_{0})/a_{0}$, with $a_{0}$ being the equilibrium lattice constant. For each strain level, the defective structures are fully relaxed with residual forces smaller than $10^{-4} \mathrm{Ry\,/\,a.u.}$. In order to study the effect of strain on the excited state properties, I calculated the ZPL for each of the defects under different levels of strain using the constrained-occupation DFT (CDFT) method.  Within the  CDFT method, the occupation of the defect states is constrained to mimic the photoluminescence process. This is done by promoting the electron from the lower-energy defect state involved in the excitation to the higher-energy defect state, which was previously unoccupied. The structure is then allowed to relax with this new electronic configuration.  The total energy differences between different electronic and ionic configurations are then mapped onto the Franck-Condon picture, giving optical transition energies.~\cite{Gali_DeltaSCF} Within the Franck-Condon picture, it is assumed that the electronic transitions upon absorption or emission of a photon happen much faster as compared to the rearrangement of the nuclei. This is illustrated in Fig.\ref{Fig1}(c), wherein, the vertical transition $\mathrm{A\rightarrow B}$ corresponds to excitation (vertical absorption), $\mathrm{C\leftrightarrow A}$ correspond to the ZPL, and $\mathrm{C\rightarrow D}$ corresponds to the vertical electronic transition with an emission of a photon.

\vspace{-0.1in}

\begin{acknowledgments}
PD thanks P. Zhang (University at Buffalo, Buffalo, NY) for discussions. This work is supported by NSF Grant number DMR-1752840 and the STC Center for Integrated Quantum Materials under NSF Grant number DMR-1231319. PD acknowledges the computational support provided by the Extreme Science and Engineering Discovery Environment (XSEDE) under Project PHY180014, which is supported by National Science Foundation grant number ACI-1548562. For three-dimensional visualization of crystals and volumetric data, use of VESTA 3 software is acknowledged. 


\end{acknowledgments}

\bibliography{hBN_PRL}

\begin{thebibliography}{31}
\expandafter\ifx\csname natexlab\endcsname\relax\def\natexlab#1{#1}\fi
\expandafter\ifx\csname bibnamefont\endcsname\relax
  \def\bibnamefont#1{#1}\fi
\expandafter\ifx\csname bibfnamefont\endcsname\relax
  \def\bibfnamefont#1{#1}\fi
\expandafter\ifx\csname citenamefont\endcsname\relax
  \def\citenamefont#1{#1}\fi
\expandafter\ifx\csname url\endcsname\relax
  \def\url#1{\texttt{#1}}\fi
\expandafter\ifx\csname urlprefix\endcsname\relax\def\urlprefix{URL }\fi
\providecommand{\bibinfo}[2]{#2}
\providecommand{\eprint}[2][]{\url{#2}}

\bibitem[{\citenamefont{Srivastava et~al.}(2015)\citenamefont{Srivastava,
  Sidler, Allain, Lembke, Kis, and Imamo\u{g}lu}}]{Imamoglu_WSe2_QE_2015}
\bibinfo{author}{\bibfnamefont{A.}~\bibnamefont{Srivastava}},
  \bibinfo{author}{\bibfnamefont{M.}~\bibnamefont{Sidler}},
  \bibinfo{author}{\bibfnamefont{A.~V.} \bibnamefont{Allain}},
  \bibinfo{author}{\bibfnamefont{D.~S.} \bibnamefont{Lembke}},
  \bibinfo{author}{\bibfnamefont{A.}~\bibnamefont{Kis}}, \bibnamefont{and}
  \bibinfo{author}{\bibfnamefont{A.}~\bibnamefont{Imamo\u{g}lu}},
  \bibinfo{journal}{Nat. Nanotechnol.} \textbf{\bibinfo{volume}{10}},
  \bibinfo{pages}{491} (\bibinfo{year}{2015}),
  \eprint{https://doi.org/10.1038/nnano.2015.60}.

\bibitem[{\citenamefont{He et~al.}(2015)\citenamefont{He, Clark, Schaibley, He,
  Chen, Wei, Ding, Zhang, Yao, Xu et~al.}}]{Pan_WSe2_QE_2015}
\bibinfo{author}{\bibfnamefont{Y.-M.} \bibnamefont{He}},
  \bibinfo{author}{\bibfnamefont{G.}~\bibnamefont{Clark}},
  \bibinfo{author}{\bibfnamefont{J.~R.} \bibnamefont{Schaibley}},
  \bibinfo{author}{\bibfnamefont{Y.}~\bibnamefont{He}},
  \bibinfo{author}{\bibfnamefont{M.-C.} \bibnamefont{Chen}},
  \bibinfo{author}{\bibfnamefont{Y.-J.} \bibnamefont{Wei}},
  \bibinfo{author}{\bibfnamefont{X.}~\bibnamefont{Ding}},
  \bibinfo{author}{\bibfnamefont{Q.}~\bibnamefont{Zhang}},
  \bibinfo{author}{\bibfnamefont{W.}~\bibnamefont{Yao}},
  \bibinfo{author}{\bibfnamefont{X.}~\bibnamefont{Xu}}, \bibnamefont{et~al.},
  \bibinfo{journal}{Nat. Nanotechnol.} \textbf{\bibinfo{volume}{10}},
  \bibinfo{pages}{497} (\bibinfo{year}{2015}),
  \eprint{https://doi.org/10.1038/nnano.2015.75}.

\bibitem[{\citenamefont{Chakraborty et~al.}(2015)\citenamefont{Chakraborty,
  Kinnischtzke, Goodfellow, Beams, and Vamivakas}}]{Vamivakas_WSe2_QE_2015}
\bibinfo{author}{\bibfnamefont{C.}~\bibnamefont{Chakraborty}},
  \bibinfo{author}{\bibfnamefont{L.}~\bibnamefont{Kinnischtzke}},
  \bibinfo{author}{\bibfnamefont{K.~M.} \bibnamefont{Goodfellow}},
  \bibinfo{author}{\bibfnamefont{R.}~\bibnamefont{Beams}}, \bibnamefont{and}
  \bibinfo{author}{\bibfnamefont{A.~N.} \bibnamefont{Vamivakas}},
  \bibinfo{journal}{Nat. Nanotechnol.} \textbf{\bibinfo{volume}{10}},
  \bibinfo{pages}{507} (\bibinfo{year}{2015}),
  \eprint{https://doi.org/10.1038/nnano.2015.79}.

\bibitem[{\citenamefont{Koperski et~al.}(2015)\citenamefont{Koperski,
  Nogajewski, Arora, Cherkez, Mallet, Veuillen, Marcus, Kossacki, and
  Potemski}}]{Potemski_WSe2_QE_2015}
\bibinfo{author}{\bibfnamefont{M.}~\bibnamefont{Koperski}},
  \bibinfo{author}{\bibfnamefont{K.}~\bibnamefont{Nogajewski}},
  \bibinfo{author}{\bibfnamefont{A.}~\bibnamefont{Arora}},
  \bibinfo{author}{\bibfnamefont{V.}~\bibnamefont{Cherkez}},
  \bibinfo{author}{\bibfnamefont{P.}~\bibnamefont{Mallet}},
  \bibinfo{author}{\bibfnamefont{J.-Y.} \bibnamefont{Veuillen}},
  \bibinfo{author}{\bibfnamefont{J.}~\bibnamefont{Marcus}},
  \bibinfo{author}{\bibfnamefont{P.}~\bibnamefont{Kossacki}}, \bibnamefont{and}
  \bibinfo{author}{\bibfnamefont{M.}~\bibnamefont{Potemski}},
  \bibinfo{journal}{Nat. Nanotechnol.} \textbf{\bibinfo{volume}{10}},
  \bibinfo{pages}{503} (\bibinfo{year}{2015}),
  \eprint{https://doi.org/10.1038/nnano.2015.67}.

\bibitem[{\citenamefont{Tran et~al.}(2015)\citenamefont{Tran, Bray, Ford, Toth,
  and Aharonovich}}]{Tran1_Nature_2015}
\bibinfo{author}{\bibfnamefont{T.~T.} \bibnamefont{Tran}},
  \bibinfo{author}{\bibfnamefont{K.}~\bibnamefont{Bray}},
  \bibinfo{author}{\bibfnamefont{M.~J.} \bibnamefont{Ford}},
  \bibinfo{author}{\bibfnamefont{M.}~\bibnamefont{Toth}}, \bibnamefont{and}
  \bibinfo{author}{\bibfnamefont{I.}~\bibnamefont{Aharonovich}},
  \bibinfo{journal}{Nat. Nanotechnol.} \textbf{\bibinfo{volume}{11}},
  \bibinfo{pages}{37} (\bibinfo{year}{2015}).

\bibitem[{\citenamefont{Tran et~al.}(2016)\citenamefont{Tran, Elbadawi,
  Totonjian, Lobo, Grosso, Moon, Englund, Ford, Aharonovich, and
  Toth}}]{Tran2_ACS_Nano2016}
\bibinfo{author}{\bibfnamefont{T.~T.} \bibnamefont{Tran}},
  \bibinfo{author}{\bibfnamefont{C.}~\bibnamefont{Elbadawi}},
  \bibinfo{author}{\bibfnamefont{D.}~\bibnamefont{Totonjian}},
  \bibinfo{author}{\bibfnamefont{C.~J.} \bibnamefont{Lobo}},
  \bibinfo{author}{\bibfnamefont{G.}~\bibnamefont{Grosso}},
  \bibinfo{author}{\bibfnamefont{H.}~\bibnamefont{Moon}},
  \bibinfo{author}{\bibfnamefont{D.~R.} \bibnamefont{Englund}},
  \bibinfo{author}{\bibfnamefont{M.~J.} \bibnamefont{Ford}},
  \bibinfo{author}{\bibfnamefont{I.}~\bibnamefont{Aharonovich}},
  \bibnamefont{and} \bibinfo{author}{\bibfnamefont{M.}~\bibnamefont{Toth}},
  \bibinfo{journal}{ACS Nano} \textbf{\bibinfo{volume}{10}},
  \bibinfo{pages}{7331} (\bibinfo{year}{2016}).

\bibitem[{\citenamefont{Choi et~al.}(2016)\citenamefont{Choi, Tran, Elbadawi,
  Lobo, Wang, Juodkazis, Seniutinas, Toth, and
  Aharonovich}}]{Aharonovich_hBN_VNCB_2016}
\bibinfo{author}{\bibfnamefont{S.}~\bibnamefont{Choi}},
  \bibinfo{author}{\bibfnamefont{T.~T.} \bibnamefont{Tran}},
  \bibinfo{author}{\bibfnamefont{C.}~\bibnamefont{Elbadawi}},
  \bibinfo{author}{\bibfnamefont{C.}~\bibnamefont{Lobo}},
  \bibinfo{author}{\bibfnamefont{X.}~\bibnamefont{Wang}},
  \bibinfo{author}{\bibfnamefont{S.}~\bibnamefont{Juodkazis}},
  \bibinfo{author}{\bibfnamefont{G.}~\bibnamefont{Seniutinas}},
  \bibinfo{author}{\bibfnamefont{M.}~\bibnamefont{Toth}}, \bibnamefont{and}
  \bibinfo{author}{\bibfnamefont{I.}~\bibnamefont{Aharonovich}},
  \bibinfo{journal}{ACS Applied Materials \& Interfaces}
  \textbf{\bibinfo{volume}{8}}, \bibinfo{pages}{29642} (\bibinfo{year}{2016}),
  \urlprefix\url{https://doi.org/10.1021/acsami.6b09875}.

\bibitem[{\citenamefont{Mart\'{\i}nez et~al.}(2016)\citenamefont{Mart\'{\i}nez,
  Pelini, Waselowski, Maze, Gil, Cassabois, and
  Jacques}}]{Single_Photon_hBN_2016}
\bibinfo{author}{\bibfnamefont{L.~J.} \bibnamefont{Mart\'{\i}nez}},
  \bibinfo{author}{\bibfnamefont{T.}~\bibnamefont{Pelini}},
  \bibinfo{author}{\bibfnamefont{V.}~\bibnamefont{Waselowski}},
  \bibinfo{author}{\bibfnamefont{J.~R.} \bibnamefont{Maze}},
  \bibinfo{author}{\bibfnamefont{B.}~\bibnamefont{Gil}},
  \bibinfo{author}{\bibfnamefont{G.}~\bibnamefont{Cassabois}},
  \bibnamefont{and} \bibinfo{author}{\bibfnamefont{V.}~\bibnamefont{Jacques}},
  \bibinfo{journal}{Phys. Rev. B} \textbf{\bibinfo{volume}{94}},
  \bibinfo{pages}{121405} (\bibinfo{year}{2016}),
  \urlprefix\url{https://link.aps.org/doi/10.1103/PhysRevB.94.121405}.

\bibitem[{\citenamefont{Grosso et~al.}(2017)\citenamefont{Grosso, Moon,
  Lienhard, Ali, Efetov, Furchi, Jarillo-Herrero, Ford, Aharonovich, and
  Englund}}]{Englund_strainTunable_Nature_2017}
\bibinfo{author}{\bibfnamefont{G.}~\bibnamefont{Grosso}},
  \bibinfo{author}{\bibfnamefont{H.}~\bibnamefont{Moon}},
  \bibinfo{author}{\bibfnamefont{B.}~\bibnamefont{Lienhard}},
  \bibinfo{author}{\bibfnamefont{S.}~\bibnamefont{Ali}},
  \bibinfo{author}{\bibfnamefont{D.~K.} \bibnamefont{Efetov}},
  \bibinfo{author}{\bibfnamefont{M.~M.} \bibnamefont{Furchi}},
  \bibinfo{author}{\bibfnamefont{P.}~\bibnamefont{Jarillo-Herrero}},
  \bibinfo{author}{\bibfnamefont{M.~J.} \bibnamefont{Ford}},
  \bibinfo{author}{\bibfnamefont{I.}~\bibnamefont{Aharonovich}},
  \bibnamefont{and} \bibinfo{author}{\bibfnamefont{D.}~\bibnamefont{Englund}},
  \bibinfo{journal}{Nat. Commun.} \textbf{\bibinfo{volume}{8}},
  \bibinfo{pages}{705} (\bibinfo{year}{2017}).

\bibitem[{\citenamefont{Palacios-Berraquero
  et~al.}(2017)\citenamefont{Palacios-Berraquero, Kara, Montblanch, Barbone,
  Latawiec, Yoon, Ott, Loncar, Ferrari, and
  Atat\"{u}re}}]{Loncar_TMD_Nature_2017}
\bibinfo{author}{\bibfnamefont{C.}~\bibnamefont{Palacios-Berraquero}},
  \bibinfo{author}{\bibfnamefont{D.~M.} \bibnamefont{Kara}},
  \bibinfo{author}{\bibfnamefont{A.~R.-P.} \bibnamefont{Montblanch}},
  \bibinfo{author}{\bibfnamefont{M.}~\bibnamefont{Barbone}},
  \bibinfo{author}{\bibfnamefont{P.}~\bibnamefont{Latawiec}},
  \bibinfo{author}{\bibfnamefont{D.}~\bibnamefont{Yoon}},
  \bibinfo{author}{\bibfnamefont{A.~K.} \bibnamefont{Ott}},
  \bibinfo{author}{\bibfnamefont{M.}~\bibnamefont{Loncar}},
  \bibinfo{author}{\bibfnamefont{A.~C.} \bibnamefont{Ferrari}},
  \bibnamefont{and}
  \bibinfo{author}{\bibfnamefont{M.}~\bibnamefont{Atat\"{u}re}},
  \bibinfo{journal}{Nat. Commun.} \textbf{\bibinfo{volume}{8}},
  \bibinfo{pages}{15093} (\bibinfo{year}{2017}),
  \eprint{https://doi.org/10.1038/ncomms15093}.

\bibitem[{\citenamefont{Exarhos et~al.}(2017)\citenamefont{Exarhos, Hopper,
  Grote, Alkauskas, and Bassett}}]{LeeBassett_2017}
\bibinfo{author}{\bibfnamefont{A.~L.} \bibnamefont{Exarhos}},
  \bibinfo{author}{\bibfnamefont{D.~A.} \bibnamefont{Hopper}},
  \bibinfo{author}{\bibfnamefont{R.~R.} \bibnamefont{Grote}},
  \bibinfo{author}{\bibfnamefont{A.}~\bibnamefont{Alkauskas}},
  \bibnamefont{and} \bibinfo{author}{\bibfnamefont{L.~C.}
  \bibnamefont{Bassett}}, \bibinfo{journal}{ACS Nano}
  \textbf{\bibinfo{volume}{11}}, \bibinfo{pages}{3328} (\bibinfo{year}{2017}),
  \bibinfo{note}{pMID: 28267917},
  \urlprefix\url{https://doi.org/10.1021/acsnano.7b00665}.

\bibitem[{\citenamefont{Tran et~al.}(2018)\citenamefont{Tran, Kianinia, Nguyen,
  Kim, Xu, Kubanek, Toth, and Aharonovich}}]{Tran3_acsphotonics_2018}
\bibinfo{author}{\bibfnamefont{T.~T.} \bibnamefont{Tran}},
  \bibinfo{author}{\bibfnamefont{M.}~\bibnamefont{Kianinia}},
  \bibinfo{author}{\bibfnamefont{M.}~\bibnamefont{Nguyen}},
  \bibinfo{author}{\bibfnamefont{S.}~\bibnamefont{Kim}},
  \bibinfo{author}{\bibfnamefont{Z.-Q.} \bibnamefont{Xu}},
  \bibinfo{author}{\bibfnamefont{A.}~\bibnamefont{Kubanek}},
  \bibinfo{author}{\bibfnamefont{M.}~\bibnamefont{Toth}}, \bibnamefont{and}
  \bibinfo{author}{\bibfnamefont{I.}~\bibnamefont{Aharonovich}},
  \bibinfo{journal}{ACS Photonics} \textbf{\bibinfo{volume}{5}},
  \bibinfo{pages}{295} (\bibinfo{year}{2018}),
  \urlprefix\url{https://doi.org/10.1021/acsphotonics.7b00977}.

\bibitem[{\citenamefont{Cress et~al.}(2016)\citenamefont{Cress, Schmucker,
  Friedman, Dev, Culbertson, Lyding, and Robinson}}]{Cress_ion_implant_2016}
\bibinfo{author}{\bibfnamefont{C.~D.} \bibnamefont{Cress}},
  \bibinfo{author}{\bibfnamefont{S.~W.} \bibnamefont{Schmucker}},
  \bibinfo{author}{\bibfnamefont{A.~L.} \bibnamefont{Friedman}},
  \bibinfo{author}{\bibfnamefont{P.}~\bibnamefont{Dev}},
  \bibinfo{author}{\bibfnamefont{J.~C.} \bibnamefont{Culbertson}},
  \bibinfo{author}{\bibfnamefont{J.~W.} \bibnamefont{Lyding}},
  \bibnamefont{and} \bibinfo{author}{\bibfnamefont{J.~T.}
  \bibnamefont{Robinson}}, \bibinfo{journal}{ACS Nano}
  \textbf{\bibinfo{volume}{10}}, \bibinfo{pages}{3714} (\bibinfo{year}{2016}),
  \eprint{https://doi.org/10.1021/acsnano.6b00252}.

\bibitem[{\citenamefont{Dean et~al.}(2010)\citenamefont{Dean, Young, Meric,
  Lee, Wang, Sorgenfrei, Watanabe, Taniguchi, Kim, Shepard
  et~al.}}]{Dean_BN_Gr_hetero_2010}
\bibinfo{author}{\bibfnamefont{C.~R.} \bibnamefont{Dean}},
  \bibinfo{author}{\bibfnamefont{A.~F.} \bibnamefont{Young}},
  \bibinfo{author}{\bibfnamefont{I.}~\bibnamefont{Meric}},
  \bibinfo{author}{\bibfnamefont{C.}~\bibnamefont{Lee}},
  \bibinfo{author}{\bibfnamefont{L.}~\bibnamefont{Wang}},
  \bibinfo{author}{\bibfnamefont{S.}~\bibnamefont{Sorgenfrei}},
  \bibinfo{author}{\bibfnamefont{K.}~\bibnamefont{Watanabe}},
  \bibinfo{author}{\bibfnamefont{T.}~\bibnamefont{Taniguchi}},
  \bibinfo{author}{\bibfnamefont{P.}~\bibnamefont{Kim}},
  \bibinfo{author}{\bibfnamefont{K.~L.} \bibnamefont{Shepard}},
  \bibnamefont{et~al.}, \bibinfo{journal}{Nat. Nanotechnol.}
  \textbf{\bibinfo{volume}{5}}, \bibinfo{pages}{722} (\bibinfo{year}{2010}),
  \eprint{https://doi.org/10.1038/nnano.2010.172}.

\bibitem[{\citenamefont{Geim and Grigorieva}(2013)}]{Geim_2D_hetero_2013}
\bibinfo{author}{\bibfnamefont{A.~K.} \bibnamefont{Geim}} \bibnamefont{and}
  \bibinfo{author}{\bibfnamefont{I.~V.} \bibnamefont{Grigorieva}},
  \bibinfo{journal}{Nature} \textbf{\bibinfo{volume}{499}},
  \bibinfo{pages}{419} (\bibinfo{year}{2013}),
  \eprint{https://doi.org/10.1038/nature12385}.

\bibitem[{\citenamefont{Cui et~al.}(2015)\citenamefont{Cui, , Lee, Kim, Arefe,
  Huang, Lee, Chenet, Zhang, Wang et~al.}}]{Hone_hetero_2015}
\bibinfo{author}{\bibfnamefont{X.}~\bibnamefont{Cui}}, ,
  \bibinfo{author}{\bibfnamefont{G.-H.} \bibnamefont{Lee}},
  \bibinfo{author}{\bibfnamefont{Y.~D.} \bibnamefont{Kim}},
  \bibinfo{author}{\bibfnamefont{G.}~\bibnamefont{Arefe}},
  \bibinfo{author}{\bibfnamefont{P.~Y.} \bibnamefont{Huang}},
  \bibinfo{author}{\bibfnamefont{C.-H.} \bibnamefont{Lee}},
  \bibinfo{author}{\bibfnamefont{D.~A.} \bibnamefont{Chenet}},
  \bibinfo{author}{\bibfnamefont{X.}~\bibnamefont{Zhang}},
  \bibinfo{author}{\bibfnamefont{L.}~\bibnamefont{Wang}}, \bibnamefont{et~al.},
  \bibinfo{journal}{Nat. Nanotechnol.} \textbf{\bibinfo{volume}{10}},
  \bibinfo{pages}{534} (\bibinfo{year}{2015}),
  \eprint{https://doi.org/10.1038/nnano.2015.70}.

\bibitem[{\citenamefont{Shotan et~al.}(2016)\citenamefont{Shotan, Jayakumar,
  Considine, Mackoit, Fedder, Wrachtrup, Alkauskas, Doherty, Menon, and
  Meriles}}]{Menon_ACS_Photonics_2016}
\bibinfo{author}{\bibfnamefont{Z.}~\bibnamefont{Shotan}},
  \bibinfo{author}{\bibfnamefont{H.}~\bibnamefont{Jayakumar}},
  \bibinfo{author}{\bibfnamefont{C.~R.} \bibnamefont{Considine}},
  \bibinfo{author}{\bibfnamefont{M.}~\bibnamefont{Mackoit}},
  \bibinfo{author}{\bibfnamefont{H.}~\bibnamefont{Fedder}},
  \bibinfo{author}{\bibfnamefont{J.}~\bibnamefont{Wrachtrup}},
  \bibinfo{author}{\bibfnamefont{A.}~\bibnamefont{Alkauskas}},
  \bibinfo{author}{\bibfnamefont{M.~W.} \bibnamefont{Doherty}},
  \bibinfo{author}{\bibfnamefont{V.~M.} \bibnamefont{Menon}}, \bibnamefont{and}
  \bibinfo{author}{\bibfnamefont{C.~A.} \bibnamefont{Meriles}},
  \bibinfo{journal}{ACS Photonics} \textbf{\bibinfo{volume}{3}},
  \bibinfo{pages}{2490} (\bibinfo{year}{2016}),
  \urlprefix\url{https://doi.org/10.1021/acsphotonics.6b00736}.

\bibitem[{\citenamefont{Jungwirth et~al.}(2016)\citenamefont{Jungwirth,
  Calderon, Ji, Spencer, Flatt\'{e}, and Fuchs}}]{Jungwirth_ACS_NanoLett_2016}
\bibinfo{author}{\bibfnamefont{N.~R.} \bibnamefont{Jungwirth}},
  \bibinfo{author}{\bibfnamefont{B.}~\bibnamefont{Calderon}},
  \bibinfo{author}{\bibfnamefont{Y.}~\bibnamefont{Ji}},
  \bibinfo{author}{\bibfnamefont{M.~G.} \bibnamefont{Spencer}},
  \bibinfo{author}{\bibfnamefont{M.~E.} \bibnamefont{Flatt\'{e}}},
  \bibnamefont{and} \bibinfo{author}{\bibfnamefont{G.~D.} \bibnamefont{Fuchs}},
  \bibinfo{journal}{Nano Letters} \textbf{\bibinfo{volume}{16}},
  \bibinfo{pages}{6052} (\bibinfo{year}{2016}), \bibinfo{note}{pMID: 27580074},
  \urlprefix\url{https://doi.org/10.1021/acs.nanolett.6b01987}.

\bibitem[{\citenamefont{Vogl et~al.}(2019)\citenamefont{Vogl, Doherty, Buchler,
  Lu, and Lam}}]{Vogl_Doherty_2019}
\bibinfo{author}{\bibfnamefont{T.}~\bibnamefont{Vogl}},
  \bibinfo{author}{\bibfnamefont{M.~W.} \bibnamefont{Doherty}},
  \bibinfo{author}{\bibfnamefont{B.~C.} \bibnamefont{Buchler}},
  \bibinfo{author}{\bibfnamefont{Y.}~\bibnamefont{Lu}}, \bibnamefont{and}
  \bibinfo{author}{\bibfnamefont{P.~K.} \bibnamefont{Lam}},
  \bibinfo{journal}{Nanoscale} \textbf{\bibinfo{volume}{11}},
  \bibinfo{pages}{14362} (\bibinfo{year}{2019}),
  \urlprefix\url{http://dx.doi.org/10.1039/C9NR04269E}.

\bibitem[{\citenamefont{Vogl et~al.}(2018)\citenamefont{Vogl, Campbell,
  Buchler, Lu, and Lam}}]{Vogl_Campbell_acsphotonics_2018}
\bibinfo{author}{\bibfnamefont{T.}~\bibnamefont{Vogl}},
  \bibinfo{author}{\bibfnamefont{G.}~\bibnamefont{Campbell}},
  \bibinfo{author}{\bibfnamefont{B.~C.} \bibnamefont{Buchler}},
  \bibinfo{author}{\bibfnamefont{Y.}~\bibnamefont{Lu}}, \bibnamefont{and}
  \bibinfo{author}{\bibfnamefont{P.~K.} \bibnamefont{Lam}},
  \bibinfo{journal}{ACS Photonics} \textbf{\bibinfo{volume}{5}},
  \bibinfo{pages}{2305} (\bibinfo{year}{2018}),
  \urlprefix\url{https://doi.org/10.1021/acsphotonics.8b00127}.

\bibitem[{\citenamefont{Lazi\'{c} et~al.}(2019)\citenamefont{Lazi\'{c},
  Espinha, Pinilla~Yanguas, Gibaja, Zamora, Ares, Chhowalla, Paz, Burgos,
  Hern\'{a}ndez-M\'{i}nguez et~al.}}]{Lazic_strain_hBN_2019}
\bibinfo{author}{\bibfnamefont{S.}~\bibnamefont{Lazi\'{c}}},
  \bibinfo{author}{\bibfnamefont{A.}~\bibnamefont{Espinha}},
  \bibinfo{author}{\bibfnamefont{S.}~\bibnamefont{Pinilla~Yanguas}},
  \bibinfo{author}{\bibfnamefont{C.}~\bibnamefont{Gibaja}},
  \bibinfo{author}{\bibfnamefont{F.}~\bibnamefont{Zamora}},
  \bibinfo{author}{\bibfnamefont{P.}~\bibnamefont{Ares}},
  \bibinfo{author}{\bibfnamefont{M.}~\bibnamefont{Chhowalla}},
  \bibinfo{author}{\bibfnamefont{W.~S.} \bibnamefont{Paz}},
  \bibinfo{author}{\bibfnamefont{J.~J.~P.} \bibnamefont{Burgos}},
  \bibinfo{author}{\bibfnamefont{A.}~\bibnamefont{Hern\'{a}ndez-M\'{i}nguez}},
  \bibnamefont{et~al.}, \bibinfo{journal}{Communications Physics}
  \textbf{\bibinfo{volume}{2}}, \bibinfo{pages}{113} (\bibinfo{year}{2019}),
  \urlprefix\url{https://doi.org/10.1038/s42005-019-0217-6}.

\bibitem[{\citenamefont{Giannozzi et~al.}(2009)\citenamefont{Giannozzi, Baroni,
  Bonini, Calandra, Car, Cavazzoni, Ceresoli, Chiarotti, Cococcioni, Dabo
  et~al.}}]{QE-2009}
\bibinfo{author}{\bibfnamefont{P.}~\bibnamefont{Giannozzi}},
  \bibinfo{author}{\bibfnamefont{S.}~\bibnamefont{Baroni}},
  \bibinfo{author}{\bibfnamefont{N.}~\bibnamefont{Bonini}},
  \bibinfo{author}{\bibfnamefont{M.}~\bibnamefont{Calandra}},
  \bibinfo{author}{\bibfnamefont{R.}~\bibnamefont{Car}},
  \bibinfo{author}{\bibfnamefont{C.}~\bibnamefont{Cavazzoni}},
  \bibinfo{author}{\bibfnamefont{D.}~\bibnamefont{Ceresoli}},
  \bibinfo{author}{\bibfnamefont{G.~L.} \bibnamefont{Chiarotti}},
  \bibinfo{author}{\bibfnamefont{M.}~\bibnamefont{Cococcioni}},
  \bibinfo{author}{\bibfnamefont{I.}~\bibnamefont{Dabo}}, \bibnamefont{et~al.},
  \bibinfo{journal}{Journal of Physics: Condensed Matter}
  \textbf{\bibinfo{volume}{21}}, \bibinfo{pages}{395502 (19pp)}
  (\bibinfo{year}{2009}), \urlprefix\url{http://www.quantum-espresso.org}.

\bibitem[{\citenamefont{Perdew and Wang}(1986)}]{GGA}
\bibinfo{author}{\bibfnamefont{J.~P.} \bibnamefont{Perdew}} \bibnamefont{and}
  \bibinfo{author}{\bibfnamefont{Y.}~\bibnamefont{Wang}},
  \bibinfo{journal}{Phys. Rev. B} \textbf{\bibinfo{volume}{33}},
  \bibinfo{pages}{8800} (\bibinfo{year}{1986}).

\bibitem[{\citenamefont{Perdew et~al.}(1996)\citenamefont{Perdew, Burke, and
  Ernzerhof}}]{PBE}
\bibinfo{author}{\bibfnamefont{J.~P.} \bibnamefont{Perdew}},
  \bibinfo{author}{\bibfnamefont{K.}~\bibnamefont{Burke}}, \bibnamefont{and}
  \bibinfo{author}{\bibfnamefont{M.}~\bibnamefont{Ernzerhof}},
  \bibinfo{journal}{Phys. Rev. Lett.} \textbf{\bibinfo{volume}{77}},
  \bibinfo{pages}{3865} (\bibinfo{year}{1996}).

\bibitem[{\citenamefont{Cohen et~al.}(2008)\citenamefont{Cohen, Mori-Sánchez,
  and Yang}}]{PBE_bandgap_issue}
\bibinfo{author}{\bibfnamefont{A.~J.} \bibnamefont{Cohen}},
  \bibinfo{author}{\bibfnamefont{P.}~\bibnamefont{Mori-Sánchez}},
  \bibnamefont{and} \bibinfo{author}{\bibfnamefont{W.}~\bibnamefont{Yang}},
  \bibinfo{journal}{Science} \textbf{\bibinfo{volume}{321}},
  \bibinfo{pages}{792} (\bibinfo{year}{2008}),
  \urlprefix\url{http://www.jstor.org/stable/20144547}.

\bibitem[{\citenamefont{Heyd et~al.}(2003)\citenamefont{Heyd, Scuseria, and
  Ernzerhof}}]{HSE03}
\bibinfo{author}{\bibfnamefont{J.}~\bibnamefont{Heyd}},
  \bibinfo{author}{\bibfnamefont{G.~E.} \bibnamefont{Scuseria}},
  \bibnamefont{and}
  \bibinfo{author}{\bibfnamefont{M.}~\bibnamefont{Ernzerhof}},
  \bibinfo{journal}{The Journal of Chemical Physics}
  \textbf{\bibinfo{volume}{118}}, \bibinfo{pages}{8207} (\bibinfo{year}{2003}),
  \eprint{https://doi.org/10.1063/1.1564060}.

\bibitem[{\citenamefont{Heyd et~al.}(2006)\citenamefont{Heyd, Scuseria, and
  Ernzerhof}}]{HSE06}
\bibinfo{author}{\bibfnamefont{J.}~\bibnamefont{Heyd}},
  \bibinfo{author}{\bibfnamefont{G.~E.} \bibnamefont{Scuseria}},
  \bibnamefont{and}
  \bibinfo{author}{\bibfnamefont{M.}~\bibnamefont{Ernzerhof}},
  \bibinfo{journal}{The Journal of Chemical Physics}
  \textbf{\bibinfo{volume}{124}}, \bibinfo{pages}{219906}
  (\bibinfo{year}{2006}), \eprint{https://doi.org/10.1063/1.2204597}.

\bibitem[{\citenamefont{Tawfik et~al.}(2017)\citenamefont{Tawfik, Ali, Fronzi,
  Kianinia, Tran, Stampfl, Aharonovich, Toth, and Ford}}]{Ford_Tran_2017}
\bibinfo{author}{\bibfnamefont{S.~A.} \bibnamefont{Tawfik}},
  \bibinfo{author}{\bibfnamefont{S.}~\bibnamefont{Ali}},
  \bibinfo{author}{\bibfnamefont{M.}~\bibnamefont{Fronzi}},
  \bibinfo{author}{\bibfnamefont{M.}~\bibnamefont{Kianinia}},
  \bibinfo{author}{\bibfnamefont{T.~T.} \bibnamefont{Tran}},
  \bibinfo{author}{\bibfnamefont{C.}~\bibnamefont{Stampfl}},
  \bibinfo{author}{\bibfnamefont{I.}~\bibnamefont{Aharonovich}},
  \bibinfo{author}{\bibfnamefont{M.}~\bibnamefont{Toth}}, \bibnamefont{and}
  \bibinfo{author}{\bibfnamefont{M.~J.} \bibnamefont{Ford}},
  \bibinfo{journal}{Nanoscale} \textbf{\bibinfo{volume}{9}},
  \bibinfo{pages}{13575} (\bibinfo{year}{2017}),
  \urlprefix\url{http://dx.doi.org/10.1039/C7NR04270A}.

\bibitem[{\citenamefont{Weston et~al.}(2018)\citenamefont{Weston,
  Wickramaratne, Mackoit, Alkauskas, and Van~de
  Walle}}]{Weston_PRB_hBN_QE_2018}
\bibinfo{author}{\bibfnamefont{L.}~\bibnamefont{Weston}},
  \bibinfo{author}{\bibfnamefont{D.}~\bibnamefont{Wickramaratne}},
  \bibinfo{author}{\bibfnamefont{M.}~\bibnamefont{Mackoit}},
  \bibinfo{author}{\bibfnamefont{A.}~\bibnamefont{Alkauskas}},
  \bibnamefont{and} \bibinfo{author}{\bibfnamefont{C.~G.} \bibnamefont{Van~de
  Walle}}, \bibinfo{journal}{Phys. Rev. B} \textbf{\bibinfo{volume}{97}},
  \bibinfo{pages}{214104} (\bibinfo{year}{2018}),
  \urlprefix\url{https://link.aps.org/doi/10.1103/PhysRevB.97.214104}.

\bibitem[{\citenamefont{Gali et~al.}(2009)\citenamefont{Gali, Janzen, Deak,
  Kresse, and Kaxiras}}]{Gali_DeltaSCF}
\bibinfo{author}{\bibfnamefont{A.}~\bibnamefont{Gali}},
  \bibinfo{author}{\bibfnamefont{E.}~\bibnamefont{Janzen}},
  \bibinfo{author}{\bibfnamefont{P.}~\bibnamefont{Deak}},
  \bibinfo{author}{\bibfnamefont{G.}~\bibnamefont{Kresse}}, \bibnamefont{and}
  \bibinfo{author}{\bibfnamefont{E.}~\bibnamefont{Kaxiras}},
  \bibinfo{journal}{Phys. Rev. Lett.} \textbf{\bibinfo{volume}{103}},
  \bibinfo{pages}{186404} (\bibinfo{year}{2009}).

\bibitem[{\citenamefont{Monkhorst and Pack}(1976)}]{kpoint2}
\bibinfo{author}{\bibfnamefont{H.~J.} \bibnamefont{Monkhorst}}
  \bibnamefont{and} \bibinfo{author}{\bibfnamefont{J.~D.} \bibnamefont{Pack}},
  \bibinfo{journal}{Phys. Rev. B} \textbf{\bibinfo{volume}{13}},
  \bibinfo{pages}{5188} (\bibinfo{year}{1976}).

\end{thebibliography}
\bibliographystyle{apsrev}


\end{document}